\documentstyle[amssymb,twocolumn,aps,prl,epsfig]{revtex}

\twocolumn

\begin{document}
\draft
\title{Coherent two-field spectroscopy of degenerate two-level systems.}
\author{A. Lezama\thanks{%
E-mail: alezama@fing.edu.uy}, S. Barreiro, A. Lipsich and A.M. Akulshin%
\thanks{%
Permanent address: Lebedev Physics Institute, 117924 Moscow, Leninsky pr.
53, Russia.}}
\address{Instituto de F\'{\i }sica, Facultad de Ingenier\'{\i }a. Casilla de correo\\
30. \\
11000, Montevideo, Uruguay.}
\date{\today }
\maketitle

\begin{abstract}
Spectroscopic features revealing the coherent interaction of a degenerate
two-level atomic system with two optical fields are examined. A model for
the numerical calculation of the response of a degenerate two-level system
to the action of an arbitrarily intense resonant pump field and a weak probe
in the presence of a magnetic field is presented. The model is valid for
arbitrary values of the total angular momentum of the lower and upper levels
and for any choice of the polarizations of the optical waves. Closed and
open degenerate two-level systems are considered. Predictions for probe
absorption and dispersion, field generation by four-wave-mixing, population
modulation and Zeeman optical pumping are derived. On all these observables,
sub-natural-width coherence resonances are predicted and their spectroscopic
features are discussed. Experimental spectra for probe absorption and
excited state population modulation in the D$_2$ line of Rb vapor are
presented in good agreement with the calculations.
\end{abstract}



\section{Introduction.}

When an atomic transition is driven by quasi-resonant light, after a
transient evolution, the quantum state describing the atomic system becomes
correlated in time with the exciting light field. This correlation is in
turn responsible for the coherent interaction between the atomic system and
the light. Of special interest is the case when the atomic system interacts
with two optical fields. In this case, in addition to the induced atomic
coherence at the frequencies of the two fields, the non-linearity of the
medium is responsible for atomic coherence at frequencies that result from
the linear combinations of the frequencies of the two incident optical
fields. These additional frequency components in the atomic dynamics
manifest themselves in the spectral dependence of different observables such
as absorption or new field generation.

The response of a pure two-level system, driven by a monochromatic pump wave
and tested by a weak probe with variable frequency offset with respect to
the pump, is well known \cite{BAKLANOV,HAROCHE,MOLLOW}. In the case of a
relatively weak pump field (Rabi frequency no larger than the transition
natural width) the coherent interaction of the two-level system with the two
fields is responsible for narrow features in probe absorption (occurring
when the two-level transition is open) whose width is determined by the
ground-state relaxation rate \cite{LETOKHOV,GORDON}. Although theoretical
predictions based in the simple two-level model have proven to be powerful
for the interpretation of a large number of experimental results, pure
two-level systems are seldom found in real experiments. In most cases, the
atomic levels are degenerate and the vectorial nature of the electromagnetic
field plays an essential role.

Well before the advent of the laser, it was realized that resonant light can
modify the state of a degenerate atomic level through optical pumping among
Zeeman sublevels\cite{HAPPER}. Experiments using optical and/or
radiofrequency excitation demonstrated the occurrence of orientation,
alignment and Zeeman coherence. The invention of the laser opened the way to
the study of the coherent response of degenerate system in the optical
domain. Essential theoretical contributions in this subject are due to
Berman and co-workers\cite{BERMAN-STEEL,BERMAN,BERMAN-GUO}. In their work,
the crucial role of optical pumping in degenerate two-level system is
underlined and it is demonstrated that sub-natural-width spectroscopic
resonances are associated to the non conservation of population, orientation
or alignment. The connection between these narrow spectroscopic features and
sub-Doppler cooling was established in \cite{BERMAN} and the pump-probe
spectroscopy of degenerate two-levels systems including the effect of the
atomic recoil was considered in \cite{BERMAN-GUO}. Bo Gao\cite{GAO,GAO-PROBE}
has developed a method allowing the calculation of the weak probe absorption
by a closed degenerate two-level system driven by a linearly polarized pump
wave and no magnetic field. The resonance fluorescence spectrum of a driven
degenerate two-level systems was qualitatively discussed with the help of
the dressed atom model\cite{REYNAUD} and explicitly derived\cite{GAO-FLUOR}.
Also, let us mention that several authors have been concerned with the
steady-state preparation of a degenerate two level system due to the
excitation by a unique field of arbitrary elliptic polarization\cite
{MORRIS,KANORSKY,GAO,MILNER,TAICHENACHEV,PRIOR}.

In this paper we are concerned with the spectroscopic features of the
coherent response of driven degenerate two level systems. Our attention is
placed on the sub-natural-width resonances associated to the coherent nature
of the interaction of an atomic system with a drive and a probe field. As
discussed below, such resonances are present in a large variety of physical
observables such as probe absorption, dispersion, fluorescence,
four-wave-mixing (FWM) etc. The aim of this paper is to discuss the
essential spectroscopic features of the coherence resonances present on
these observables and discuss the dependence of these features in parameters
such as external magnetic field, optical fields polarizations, light
intensity and atomic level characteristics.

Additional motivation for the study of degenerate two-level systems arise
from the connection of this problem with that of the study of multilevel
configurations where interesting coherent effects have attracted
considerable attention in recent years. Among them is the phenomenon of
coherent population trapping (CPT)\cite{ARIMONDO} and the related effect of
electromagnetic induced transparency (EIT)\cite{HARRIS} observed in
three-level systems (mainly $\Lambda $ systems). These effects have found
interesting application for subrecoil laser cooling\cite{LAWALL},
magnetometry\cite{LEE}, refractive index enhancement\cite{REFRACT},
enhancement of non-linear susceptibility\cite{SUSCEPT}, steep dispersion\cite
{STEEP} and ultra-low group velocity propagation\cite{HAU}. Degenerate
two-level systems provide us with the possibility to analyze some of the
level schemes studied so far within a unique theoretical frame. Under the
appropriate choice of the pump and probe polarizations and the angular
momenta of the involved levels, the degenerate two level system in the
presence of two-fields reduces, as a particular case, into several of the
multi-level configurations previously studied ($\Lambda ,V,N$
configurations, etc.). One can thus expect that degenerate two level systems
will be suitable for the observation of coherent effects including those
previously predicted and observed in three level configurations. In
addition, new effects such as Electromagnetically Induced Absorption (EIA)%
\cite{LEZAMA,LEZAMA2} appear in the degenerate two level system that are not
present in three-level configurations.

The first part of the paper is devoted to the presentation of a theoretical
model allowing the calculation of the complete response to first order in
the probe field of an atomic system composed of two degenerate levels of
given total angular momentum driven by a pump wave. The model is based in a
semiclassical treatment of the atom+fields dynamics based in optical Bloch
equations\cite{BERMAN-STEEL,BERMAN,BERMAN-GUO,GAO,GAO-PROBE}. It is intended
to be suitable for the numerical calculation of the response of the atomic
medium in a wide variety of cases: Arbitrary values of the total angular
momentum of the ground and excited levels can be considered. Arbitrary and
independent elliptical polarizations are allowed for the pump and probe
waves. The presence of an external magnetic field is included. Both open and
closed two-level systems can be treated.

The model results in the derivation of the equation \ref{eqprobe} below
satisfied by the operator $\sigma $ defined in order to contain all density
matrix terms corresponding to the response of the atomic system (under the
presence of the arbitrarily intense pump wave) to first order in the probe
wave. Different spectroscopic observables such as probe absorption and
dispersion, new field generation, population modulation, magnetic
orientation, are subsequently derived from the operator $\sigma $ as
discussed below. The spectral features present on these observables are then
discussed with special attention on the subnatural-width resonances due to
the coherent interaction between the atomic system and the fields. The
predictions are illustrated with experimentally observations of probe
absorption and fluorescence modulation carried on the D$_2$ lines of Rb
vapor.

The paper is organized as follows. The second section is devoted to the
presentation of the theoretical model and to the derivation of the
expressions for different observables. The spectroscopic features of each of
these observables as a function of the optical and magnetic field parameters
are specifically discussed. The third section is devoted to experimental
observations concerning probe absorption and population modulation spectra.
Finally, some conclusions are presented.

\section{Theory.}

\subsection{Model.}

\begin{figure}[tbp]
\begin{center}
\mbox{\epsfig{file=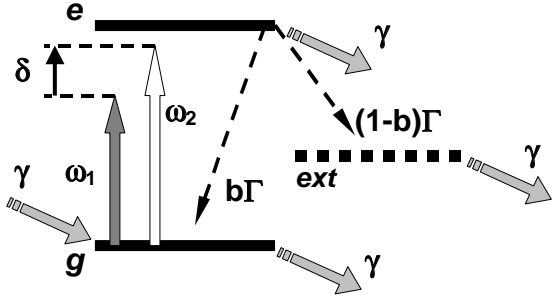,width=3.5in}}
\end{center}
\caption{Level scheme considered for the model. The ground level $g$ and the
excited level $e$ are degenerate with total angular momentum $F_g$ and $F_e$
respectively. Vertical arrows: excitation fields. Black dashed arrows:
radiative decay. Oblique gray arrows: arrival and departure from the
interaction zone.}
\label{lev_scheme}
\end{figure}

The model was developed having in mind transitions within the D lines of
alkaline atoms. However, it is not restricted to this case. It can be
applied to any dipole allowed atomic transition from a long-living lower
level to an upper level rapidly damped by spontaneous emission. We consider
two degenerate levels: a ground level $g$ of total angular momentum $F_g$
and energy $\hbar \omega _g$ and an excited state $e$ of angular momentum $%
F_e$ and energy $\hbar \omega _e$. The total radiative relaxation
coefficient of level $e$ is $\Gamma $. We assume that the atoms in the
excited state can radiatively decay into the ground state $g\ $at a the rate 
$b\Gamma $, where $b$ is a branching ratio coefficient that depends on the
specific atomic transition $\left( 0\leq b\leq 1\right) $. $b=1$ for a
closed (cycling) transition. In the case of open transitions $(b<1)$,
excited atoms can decay back into level $g$ or into one or several levels
external to the two-level system where they remain (see Fig. \ref{lev_scheme}%
).

Let consider a homogeneous ensemble of atoms at rest. However, in order to
simulate the effect of a finite interaction time of the atoms with the
light, we assume that the atoms escape the interaction region at a rate $%
\gamma $ , $(\gamma \ll \Gamma )$. This escape is compensated, at
steady-state, by the arrival of fresh atoms in the ground state.

The atoms are submitted to the action of a magnetic field $B$ and two
classical optical fields:

\begin{quotation}
\begin{equation}
\vec{E}_j(t)=E_j\hat{e}_je^{i\omega _jt}+E_j^{*}\hat{e}_j^{*}e^{-i\omega
_jt},\ (j=1,2)  \label{flds}
\end{equation}

$\widehat{e}_j$ are complex polarization vectors. The Hamiltonian of the
system can be written as:
\end{quotation}

\begin{equation}
H(t)=H_0+V_1\left( t\right) +V_2\left( t\right)  \label{hamtot}
\end{equation}
with:

\begin{eqnarray}
H_0 &=&H_A+H_B  \label{hab} \\
H_A &=&\hbar \left( P_e\omega _e+P_g\omega _g\right)  \label{ha} \\
H_B &=&(\beta _gP_g+\beta _eP_e)F_zB  \label{hb} \\
V_j &=&E_j\hat{e}_j\cdot \vec{D}_{ge}e^{i\omega _jt}+E_j^{*}\hat{e}%
_j^{*}\cdot \vec{D}_{eg}e^{-i\omega _jt}  \label{atomfield}
\end{eqnarray}
here $P_g$ and $P_e$ are the projectors on the ground and excited manifolds
respectively. $\beta _g$ and $\beta _e$ are the gyromagnetic factors of
levels $g$ and $e$ respectively. $F_z$ is the component of the total angular
momentum operator $\vec{F}$ along the magnetic field direction. $\vec{D}%
_{ge}=\vec{D}_{eg}^{\dagger }=P_g\vec{D}P_e$ is the lowering part of the
atomic dipole operator (we assume that $P_g\vec{D}P_g=P_e\vec{D}P_e=0$). In
Eq. \ref{atomfield} the usual rotating wave approximation is used.

The temporal evolution of the atomic density matrix $\rho $ is governed by
the master equation \cite{COHEN}:

\begin{eqnarray}
\frac{\partial \rho }{\partial t} &=&-\frac i\hbar \left[ H,\rho \right] -%
\frac \Gamma 2\left\{ P_{e,}\rho \right\}  \nonumber \\
&&+b\Gamma \sum_{q=-1,0,1}Q_{ge}^q\rho Q_{eg}^q-\gamma \left( \rho -\rho
_0\right)  \label{mastereq}
\end{eqnarray}
where $Q_{ge}^q=Q_{eg}^{q\dagger }\ \left( q=-1,0.1\right) $ are the
standard components of the dimensionless operator:

$\smallskip $%
\begin{equation}
\vec{Q}_{ge}=\sqrt{2F_e+1}\frac{\vec{D}_{ge}}{\left\langle g\Vert \vec{D}%
\Vert e\right\rangle }  \label{defq}
\end{equation}
$F_{e(g)}$ is the total angular momentum of the excited(ground) state and $%
\left\langle g\Vert \vec{D}\Vert e\right\rangle $ is the reduced matrix
element of the electric dipole operator.

The first term on the right hand side Eq. \ref{mastereq} represents the free
atomic evolution in the presence of the two optical fields and the magnetic
field. The remaining terms correspond to atomic relaxation. The second term
on the right hand side accounts for the radiative relaxation of the excited
state. The third term describes the feeding of the ground level by atoms
decaying from the excited state. The last term phenomenologically accounts
for the finite interaction time and ensures the relaxation of the system, in
the absence of optical fields, to the thermal equilibrium state assumed to
be described by the density matrix $\rho _0=P_g/(2F_g+1)$ corresponding to
an isotropic distribution of the atomic population in the ground-state.
Since there is no specific ground-state relaxation mechanism and $\gamma \ll
\Gamma $, the rate $\gamma $ effectively plays the role of a ground-state
relaxation coefficient.

In order to find the response of the system to the two fields we first
consider the effect of field $E_1$ (hereafter designated as {\it pump}
field) to all orders. Next, we calculate the effect of field $E_2$ ({\it %
probe} field) to first order. This procedure is analogous to the employed in
the classical deduction of Mollow\cite{MOLLOW}.

To obtain the response of the atomic system to the pump it is convenient to
introduce the slowly varying matrix $\sigma _0$ given by: 
\begin{eqnarray}
\sigma ^0 &=&\sigma _{gg}^0+\sigma _{ee}^0+\sigma _{ge}^0+\sigma _{eg}^0 
\nonumber \\
\sigma _{gg}^0 &=&P_g\rho P_g  \nonumber \\
\sigma _{ee}^0 &=&P_e\rho P_e  \label{sigma0} \\
\sigma _{ge}^0 &=&P_g\rho P_ee^{-i\omega _1t}  \nonumber \\
\sigma _{eg}^0 &=&P_e\rho P_ge^{i\omega _1t}  \nonumber
\end{eqnarray}
after substitution into Eq. \ref{mastereq} (with $V_2=0$) one has for the
steady state value of $\sigma ^0$ the equation: 
\begin{eqnarray}
-\gamma \rho _0 &=&-\frac i\hbar \left[ H_0+\bar{V}_1-\hbar \omega
_1P_e,\sigma ^0\right] -\frac \Gamma 2\left\{ P_e,\sigma ^0\right\} 
\nonumber \\
&&+b\Gamma \sum_{q=-1,0,1}Q_{ge}^q\sigma ^0Q_{eg}^q-\gamma \sigma ^0
\label{pumpeq}
\end{eqnarray}
\onecolumn
with: 
\begin{equation}
\bar{V}_1=E_1\hat{e}_1\cdot \vec{D}_{ge}+E_1^{*}\hat{e}_1^{*}\cdot \vec{D}%
_{eg}  \label{fieldslow}
\end{equation}

Eq. \ref{pumpeq} represents a system of coupled linear equations involving
the coefficients of the density matrix $\sigma ^0$. We have numerically
solved this system by employing a Liouville space formalism\cite
{FEUILLADE,CYR} in which the state of the system is represented by a vector $%
{\bf x}$ whose elements are the coefficients of the density matrix $\sigma
^0 $. In the Liouville space Eq. \ref{pumpeq} can be put into the form $%
{\cal L}{\bf x}={\bf x}_{{\bf 0}}$ (${\cal L}$ is a linear operator and $%
{\bf x}_{{\bf 0}}$ a constant vector) and readily inverted.

The relaxation term describing the return of the unperturbed system to
equilibrium (fourth term in the right hand side of Eq. \ref{pumpeq}), as
well as the possible escape to external levels when $b<1$, do not conserve
the total population. In consequence, the solution of Eq. \ref{pumpeq} must
be normalized by the total number of atoms $N=n_g+n_e+n_{ext}$ where $n_g$, $%
n_e$ and $n_{ext}$ are the populations of the ground level, excited level
and external level(s) respectively. Since at steady state the population of
the external level(s) obeys the balance equation: $\Gamma (1-b)n_e=\gamma
n_{ext}$ we have: 
\begin{equation}
N=n_g+n_e[1+(1-b)\Gamma /\gamma ]  \label{ntot}
\end{equation}
where $n_g$ and $n_e$ are the ground and excited-state populations obtained
from the numerical solution of \ref{pumpeq} respectively. After
normalization the solution of Eq. \ref{pumpeq}, $\sigma ^0$ fully accounts
for the preparation of the system by the pump. This includes effects such as
optical saturation and induced coherence between ground and excited states,
orientation and alignment due to Zeeman optical pumping and depopulation.

To analyze the effect to first order of the probe field on the atom + pump
system, we seek a solution of Eq. \ref{mastereq} under the form\cite{MOLLOW}%
: 
\begin{eqnarray}
\rho _{gg}(t) &=&P_g\rho \left( t\right) P_g=\sigma _{gg}^0+\sigma
_{gg}^{+}e^{i\delta t}+\sigma _{gg}^{-}e^{-i\delta t}  \nonumber \\
\rho _{ee}(t) &=&P_e\rho \left( t\right) P_e=\sigma _{ee}^0+\sigma
_{ee}^{+}e^{i\delta t}+\sigma _{ee}^{-}e^{-i\delta t}  \nonumber \\
\rho _{ge}(t) &=&P_g\rho \left( t\right) P_e=e^{i\omega _1t}\left( \sigma
_{ge}^0+\sigma _{ge}^{+}e^{i\delta t}+\sigma _{ge}^{-}e^{-i\delta t}\right)
\label{firstorder} \\
\rho _{eg}(t) &=&P_e\rho \left( t\right) P_g=e^{-i\omega _1t}\left( \sigma
_{eg}^0+\sigma _{eg}^{+}e^{i\delta t}+\sigma _{eg}^{-}e^{-i\delta t}\right) 
\nonumber
\end{eqnarray}
with $\delta =\omega _2-\omega _1$. After substitution into Eq.\ref{mastereq}
and keeping only terms to first order in $E_2$ one has the following system
of linear equations coupling $\sigma _{ge}^{+}$, $\sigma _{eg}^{+}$, $\sigma
_{gg}^{+}$ and $\sigma _{ee}^{+}$: 
\begin{eqnarray}
iE_2\widehat{(e}_2.\vec{D}_{ge})\sigma _{eg}^0 &=&-i\left[ H_g,\sigma
_{gg}^{+}\right] -i\hbar \delta \sigma _{gg}^{+}+b\hbar \Gamma
\sum_{q=-1,0,1}Q_{ge}^q\sigma _{ee}^{+}Q_{eg}^q  \nonumber \\
&&-\hbar \gamma \sigma _{gg}^{+}-i\left( E_1\hat{e}_1\cdot \vec{D}%
_{ge}\sigma _{eg}^{+}-\sigma _{ge}^{+}E_1^{*}\hat{e}_1^{*}\cdot \vec{D}%
_{eg}\right)  \nonumber \\
-iE_2\sigma _{eg}^0\widehat{(e}_2.\vec{D}_{ge}) &=&-i\left[ H_e,\sigma
_{ee}^{+}\right] -i\hbar \delta \sigma _{ee}^{+}-\hbar \left( \Gamma +\gamma
\right) \sigma _{ee}^{+}  \nonumber \\
&&-i\left( E_1^{*}\hat{e}_1^{*}\cdot \vec{D}_{eg}\sigma _{ge}^{+}-\sigma
_{eg}^{+}E_1\hat{e}_1\cdot \vec{D}_{ge}\right)  \label{eqsyst} \\
i\left( E_2\widehat{e}_2.\vec{D}_{ge}\sigma _{ee}^0-\sigma _{gg}^0E_2%
\widehat{e}_2.\vec{D}_{ge}\right) &=&-i\left( H_g\sigma _{ge}^{+}-\sigma
_{ge}^{+}H_e\right) -i\hbar \left( \delta +\omega _1\right) \sigma
_{ge}^{+}-\hbar \left( \frac \Gamma 2+\gamma \right) \sigma _{ge}^{+} 
\nonumber \\
&&-i\left( E_1\hat{e}_1\cdot \vec{D}_{ge}\sigma _{ee}^{+}-\sigma _{gg}^{+}E_1%
\hat{e}_1\cdot \vec{D}_{ge}\right)  \nonumber \\
0 &=&-i\left( H_e\sigma _{eg}^{+}-\sigma _{eg}^{+}H_g\right) -i\hbar \left(
\delta -\omega _1\right) \sigma _{eg}^{+}-\hbar \left( \frac \Gamma 2+\gamma
\right) \sigma _{eg}^{+}  \nonumber \\
&&-i\left( E_1^{*}\hat{e}_1^{*}\cdot \vec{D}_{eg}\sigma _{gg}^{+}-\sigma
_{ee}^{+}E_1^{*}\hat{e}_1^{*}\cdot \vec{D}_{eg}\right)  \nonumber
\end{eqnarray}
It is convenient at this point to introduce the non-Hermitian matrix $\sigma 
$ defined as: 
\begin{equation}
\sigma =\left( 
\begin{array}{ll}
\sigma _{gg}^{+} & \sigma _{ge}^{+} \\ 
\sigma _{eg}^{+} & \sigma _{ee}^{+}
\end{array}
\right)  \label{matrix}
\end{equation}

After inspection it can be seen that Eqs.\ref{eqsyst} can be then rewritten
in the operatorial form:

\begin{eqnarray}
i\left[ \frac{\Omega _2}2,\sigma _0\right] &=&-\frac i\hbar \left[ H_0+\bar{V%
}_1-\hbar \omega _1P_e,\sigma \right]  \nonumber \\
&&-i\delta \sigma -\frac \Gamma 2\left\{ P_e,\sigma \right\} -\gamma \sigma 
\nonumber \\
&&+b\Gamma \sum_{q=-1,0,1}Q_{ge}^q\sigma Q_{eg}^q  \label{eqprobe}
\end{eqnarray}
\twocolumn
where we have introduced the probe Rabi frequency $\Omega _2=(2\vec{E}%
_2\cdot \vec{D}_{ge})/\hbar .$

Eq. \ref{eqprobe} is numerically solved by the same Liouville-space
procedure used to solve Eq. \ref{pumpeq}. The resulting value of the
operator $\sigma $ simultaneously provides the values of the four matrices $%
\sigma _{ge}^{+}$, $\sigma _{eg}^{+}$, $\sigma _{gg}^{+}$ and $\sigma
_{ee}^{+}$ each of which contains useful information on the atomic response
as will be now discussed.

The information concerning the optical atomic polarization oscillating at
the probe frequency is contained in matrix $\sigma _{ge}^{+}$ (see Eqs. \ref
{firstorder}). The complex macroscopic atomic polarization at the frequency
of the probe is given by: 
\begin{equation}
\vec{P}\left( \omega _2\right) =\varepsilon _0\chi \vec{E}_2=nTr\left(
\sigma _{ge}^{+}\vec{D}_{eg}\right)  \label{atompolar}
\end{equation}
where $\varepsilon _0$ is the vacuum dielectric constant, $\chi =\chi
^{\prime }+i\chi ^{\prime \prime }$ is the complex susceptibility tensor and 
$n$ is the atom density. From the complex polarization we get the probe
absorption coefficient as: 
\begin{equation}
\alpha \left( \omega _2\right) =k_2(\widehat{e}_2\cdot \chi ^{\prime \prime }%
\widehat{e}_2)=\frac{nk_2}{E_2}\widehat{e}_2\cdot Imag\left[ Tr\left( \sigma
_{ge}^{+}\vec{D}_{eg}\right) \right]  \label{abscoef}
\end{equation}
$k_2$ is the modulus of the probe wave-vector.

Similarly, the dielectric tensor $\varepsilon =\varepsilon _0(1+\chi
^{\prime })$ describing the dispersion of the probe by the atomic medium
obeys: 
\begin{equation}
\widehat{e}\cdot \varepsilon \ \widehat{e}_2=\varepsilon _0\left\{ 1+\frac n{%
E_2}\widehat{e}\cdot Real\left[ Tr\left( \sigma _{ge}^{+}\vec{D}_{eg}\right)
\right] \right\}  \label{refindex}
\end{equation}

where $\widehat{e}$ is an arbitrary unit vector.

The matrix $\sigma _{eg}^{+}$ contains information about the off-diagonal
elements of the density matrix evolving at the frequency $2\omega _1-\omega
_2$. The corresponding atomic polarization is responsible for the generation
of a new optical field at this frequency by four wave-mixing (FWM) \cite
{SHEN}.

The intensity of the field generated by FWM satisfies:

\begin{equation}
I_{FWM}(2\omega _1-\omega _2)\varpropto n^2\left| Tr\left( \sigma _{eg}^{+}%
\vec{D}_{ge}\right) \right| ^2  \label{fwm}
\end{equation}

The matrices $\sigma _{gg}^{+}$ and $\sigma _{ee}^{+}$ contain information
concerning the pulsation of the ground and excited state populations at the
difference frequency $\delta $ between the pump and the probe fields.
Opposite fluctuation of the total population occurs in the ground and the
excited state. The effect of the modulation of the population of the excited
state can be directly observed as an oscillating component (at frequency $%
\delta $) of the fluorescence emitted by the atoms. The pulsation of the
total fluorescence power emitted per atom is:

\begin{equation}
\Delta W(t)=2b\Gamma \hbar \omega _0\left[ Tr\left( \sigma _{ee}^{+}\right)
e^{i\delta t}+cc\right]  \label{acfluor}
\end{equation}

Finally, the matrices $\sigma _{gg}^{+}$ and $\sigma _{ee}^{+}$ also provide
information about population redistribution and coherence among Zeeman
sublevels (orientation and/or alignment) in the ground or excited state due
to the simultaneous interaction with the pump and probe fields. This result
in additional observables evolving at frequency $\delta $ . As an example,
the oscillating magnetic dipole induced in the ground state by the coherent
interaction with the pump and probe fields is given by:

\begin{equation}
\vec{M}_g(t)=n\beta _g\left[ Tr\left( \sigma _{gg}^{+}\vec{F}\right)
e^{i\delta t}+cc\right]  \label{mag_dip}
\end{equation}

In the following sections we present and discuss the spectroscopic features
of the atomic response. Particular attention is given to the absorption and
population pulsation spectra for which experimental results are presented.

\subsection{Probe absorption spectra.}

\subsubsection{Spectra with zero magnetic field.}

\begin{figure}[tbp]
\begin{center}
\mbox{\epsfig{file=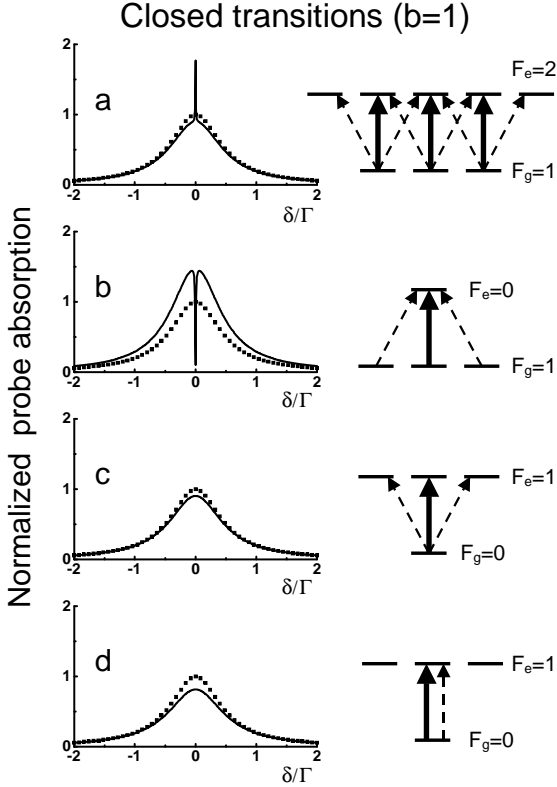,width=3.5in}}
\end{center}
\caption{Probe absorption spectra as a function of the frequency offset $%
\delta $ (solid lines) with $B=0$ and $\Omega _1/\Gamma \equiv
2E_1\left\langle g\left\| D\right\| e\right\rangle /\left( \hbar \Gamma
\right) =0.4,\ \gamma /\Gamma =0.001$ for different closed ($b=1$) atomic
transitions. The spectra are normalized to the maximum of the linear
absorption (represented in dotted lines). The pump and probe polarizations
are linear and orthogonal in $a$, $b$, $c$ and parallel in $d$. The
corresponding angular momenta and magnetic sublevels scheme is shown in each
case. Solid (dashed) arrows represent the pump (probe) field.}
\label{spect_rest_closed}
\end{figure}

\begin{figure}[tbp]
\begin{center}
\mbox{\epsfig{file=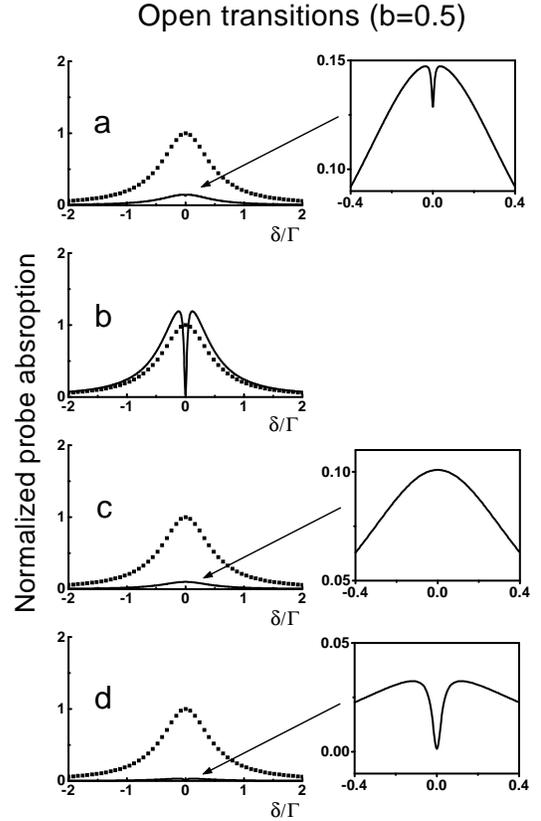,width=3.5in}}
\end{center}
\caption{Probe absorption spectra as a function of the frequency offset $%
\delta $ (solid lines) with $B=0$ and $\Omega _1/\Gamma \equiv
2E_1\left\langle g\left\| D\right\| e\right\rangle /\left( \hbar \Gamma
\right) =0.4,\ \gamma /\Gamma =0.001$ for different open ($b=0.5$) atomic
transitions. The spectra are normalized to the maximum of the linear
absorption (represented in dotted lines). The fields polarizations are the
same as in the corresponding case of Fig. \ref{spect_rest_closed}. The
insets represent enlarged views around the center of the spectra.}
\label{spect_rest_open}
\end{figure}

We initially consider the results obtained for motionless atoms in the
absence of a magnetic field. Some examples of level configurations and the
corresponding probe absorption spectra for particular choices of the
polarizations of the pump and probe fields are shown in Figs. \ref
{spect_rest_closed} and \ref{spect_rest_open}. In general, the spectra are
composed of a Lorentzian peak, whose width is given by the excited state
relaxation rate and in most cases, a narrow resonance (hereafter named
coherence resonance) arising when the frequency offset $\delta \ $between
pump and probe satisfies the condition for two-photon Raman resonances
between Zeeman sublevels. In the particular case of Figs. \ref
{spect_rest_closed} and \ref{spect_rest_open} a unique coherence resonance
at $\delta =0$ occurs since $B=0$. Let us first discuss the case of closed
(cycling) transitions. Depending on the choice of the total angular momentum
of the ground and excited states the coherence resonance corresponds to a
reduction in the absorption (EIT) or an enhancement of the absorption (EIA)%
\cite{LEZAMA,LEZAMA2}. No coherence resonance appears for some
configurations. As discussed in Ref. \cite{LEZAMA2} EIA occurs provided that
three conditions are satisfied: $i)$ The transition is closed, $ii)\ F_g>0$
and $iii)$ $F_g<F_e$. This is the case of Fig. \ref{spect_rest_closed} (a).
On the other hand, EIT occurs if $F_g\geq F_e$ and $F_g>0$ [Fig. \ref
{spect_rest_closed} (b)]\cite{SMIRNOV}. Finally, no coherence resonance
occurs if $F_g=0$ [Fig. \ref{spect_rest_closed}(c)].

When the transition is open $\left( b<1\right) $, the previous results are
modified. In addition to an overall reduction of the absorption as a
consequence of population loss to the external levels significant changes
concern the coherence resonance. When $F_g<F_e$ and $Fg>0$ for sufficiently
large values of the escape rate $(1-b)\Gamma $ compared to $\gamma $ a
narrow transparency dip appears at the place of the absorption enhancement
observed for cycling transitions\cite{GORDON} [Fig. \ref{spect_rest_open}
(a)]. A similar absorption dip is present in the case of the $%
F_g=0\rightarrow F_e=1$ transition with equal pump and probe polarizations
[Fig. \ref{spect_rest_open} (d)]. A situation that corresponds to the pure
two-level system. The origin of the narrow dip is the same in the two cases.
It can be attributed to the resonant scattering of the probe field by the
modulation in the atomic populations induced by the two fields\cite
{BAKLANOV,SHEN}. As in the case of closed transitions, no narrow resonance
occurs for the $F_g=0\rightarrow F_e=1$ transition when the polarizations
are orthogonal ($V$ system) [Fig. \ref{spect_rest_open} (c)]. Finally when $%
F_g\geq F_e$ [Fig. \ref{spect_rest_open} (b)] a large EIT resonance is
observed. As in the case of cycling transitions, the transparency dip is
essentially associated to the falling of the system into a dark state in the
ground level rather than to an increased escape to external levels\cite
{PRIOR}.

In the present context, both EIT and EIA appear in the same theoretical
framework as the result of the coherent interaction of the degenerate atomic
system with the two fields. This result clearly stresses the role of the
atomic dynamics at the common origin of the two phenomena{\it . }It is
nevertheless interesting to point out that from a different point of view,
EIT has been usually associated to the existence of a stationary {\it %
dark-state} (a quantum superposition of ground state sublevels not coupled
to the excited state) in which the atomic population is trapped \cite
{ARIMONDO}. This arises the question of whether a corresponding ``{\it %
bright-state}'' can be identified to correspond to EIA.

The identification of the steady state of the system, in the case of EIA, as
well as for EIT, is greatly simplified in the present context (degenerate
levels, $B=0$) since at resonance ($\delta =0$) the pump and probe fields
reduce to a single field. Consequently, the stationary state of the system
is simply the result of the optical pumping of the degenerate two
level-system by the unique field. When the conditions for EIT are satisfied (%
$F_g\geq F_e$) this results in the falling of the system into the uncoupled
sub-space of the ground level. When EIA occurs ($F_g<F_e$), all the ground
level manifold is affected by the light. In this case, the steady state of
the system is shared by the ground and the excited level. Unlike in the case
of EIT, the steady state corresponding to an EIA resonance is obviously not
stable (in the absence of the fields) since it is affected by spontaneous
emission.

\begin{figure}[tbp]
\begin{center}
\mbox{\epsfig{file=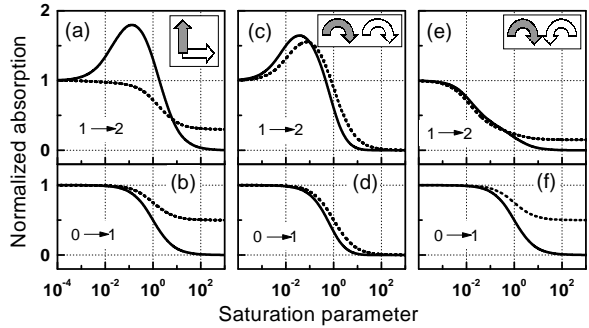,width=3.5in}}
\end{center}
\caption{Absorption at $\delta =0$, $B=0,\ \gamma /\Gamma =0.001$ for
different closed atomic transitions and different pump and probe
polarizations as a function of the saturation parameter $S$ $\equiv
2[2E_1\left\langle g\left\| D\right\| e\right\rangle /\left( \hbar \Gamma
\right) ]^2$ (solid lines). The absorption is normalized to the maximum
linear absorption. The dashed lines represent the incoherent optical pumping
contribution (see text). The corresponding $F_g\rightarrow F_e=F_g+1$
transition is indicated in each case. The pump and probe polarizations are
linear and orthogonal ($a,b$), circular and equal ($c,d$) and circular and
opposite ($e,f$).}
\label{peak-sat}
\end{figure}

We address now the question of the amplitude of the coherence resonance in
the probe absorption. We restrict ourselves to closed transitions of the
type $F_g\rightarrow F_e=F_g+1$ where EIA occurs. The coherence resonance
amplitude depends on the angular momenta of the levels considered, the
optical polarizations, the pump intensity and the rate $\gamma $. Some
examples of the intensity dependence of the absorption at $\delta =0$ for
different $F_g$ ($F_e=F_g+1$) and the optical polarizations are shown in
Fig. \ref{peak-sat}. The absorption is normalized to the probe absorption in
the absence of the pump and it is presented as a function of the saturation
parameter $S\equiv 2[2E_1\left\langle e\left\| D\right\| g\right\rangle
/(\hbar \Gamma )]^2$ which is proportional to pump intensity. Also shown in
Fig. \ref{peak-sat} is the contribution to the probe absorption which is
only due to the incoherent optical pumping by the pump field. This
contribution, obtained taking $\bar{V}_1=0$ in Eq. \ref{eqprobe}, includes
the effect of Zeeman optical pumping and optical saturation produced by the
pump field but ignores coherent two-photon processes involving both fields.
The maximum resonant enhancement of the absorption occurs for linear and
perpendicular pump and probe polarizations. Notice that in this case, the
incoherent optical pumping contribution remains below the linear absorption
for all intensities. When the polarization of the two fields are the same,
(Fig. \ref{peak-sat}c,d) the coherent enhancement of the absorption is
smaller and there is a significant contribution to the absorption
enhancement from incoherent optical pumping. No absorption enhancement
occurs for circular and opposite pump and probe polarizations Fig. \ref
{peak-sat}(e,f). Notice that Figs. \ref{peak-sat}(d) and \ref{peak-sat}(f)
correspond to the pure two-level and V schemes respectively.

\begin{figure}[tbp]
\begin{center}
\mbox{\epsfig{file=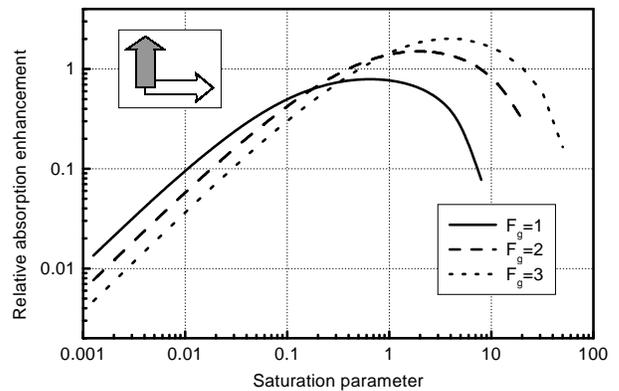,width=3.5in}}
\end{center}
\caption{Relative absorption enhancement $\eta =(\alpha -\alpha _L)/\alpha
_L\ $at $\delta =0$, $B=0,\ \gamma /\Gamma =0.001$ for different $%
F_g\rightarrow F_e=F_g+1$ closed atomic transitions as a function of the
saturation parameter $S$ ($\alpha _L$ is the maximum linear absorption). The
pump and probe polarizations are linear and orthogonal.}
\label{loglog}
\end{figure}

As is also the case for EIT\cite{SCULLY}, in the weak pump limit the
amplitude of the EIA resonance is linear in the pump intensity as shown in
Fig. \ref{loglog} for linear and perpendicular pump and probe polarizations 
\cite{TAICHENACHEV2}. Notice that for a weak pump field the higher
absorption enhancement factor is obtained for $F_g=1$ and decreases for
increasing $F_g$. However, the maximum value of the enhancement increases
with $F_g$ for the values considered in Fig. \ref{loglog}. The maximum
absorption enhancement is obtained for a saturation parameter close to
unity. At higher intensities the absorption at zero frequency offset
decreases as a consequence of the dynamic Stark effect\cite{MOLLOW}.

\begin{figure}[tbp]
\begin{center}
\mbox{\epsfig{file=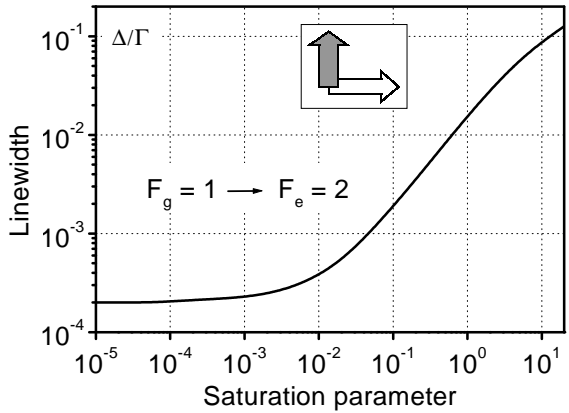,width=3.5in}}
\end{center}
\caption{Linewidth $\Delta $ (half width at half maximum) of the coherence
absorption resonance at $\delta =0$, $B=0,\ \gamma /\Gamma =10^{-4} $ for
the closed $F_g=1\rightarrow F_e=2$ transition as a function of the
saturated parameter $S$. The pump and probe polarizations are linear and
orthogonal.}
\label{width}
\end{figure}

The linewidth $\Delta $ of the EIA and EIT resonances (FWHM) is determined,
at low pump field intensities, by the finite interaction time. It is given
by $\Delta =2\gamma $. At larger pump intensities power broadening occurs
with $\Delta $ increasing linearly with the pump intensity (see Fig. \ref
{width})\cite{LEZAMA}. This behavior is analogous to the observed for EIT
resonances in $\Lambda $ configurations \cite{ARIMONDO}.

\begin{figure}[tbp]
\begin{center}
\mbox{\epsfig{file=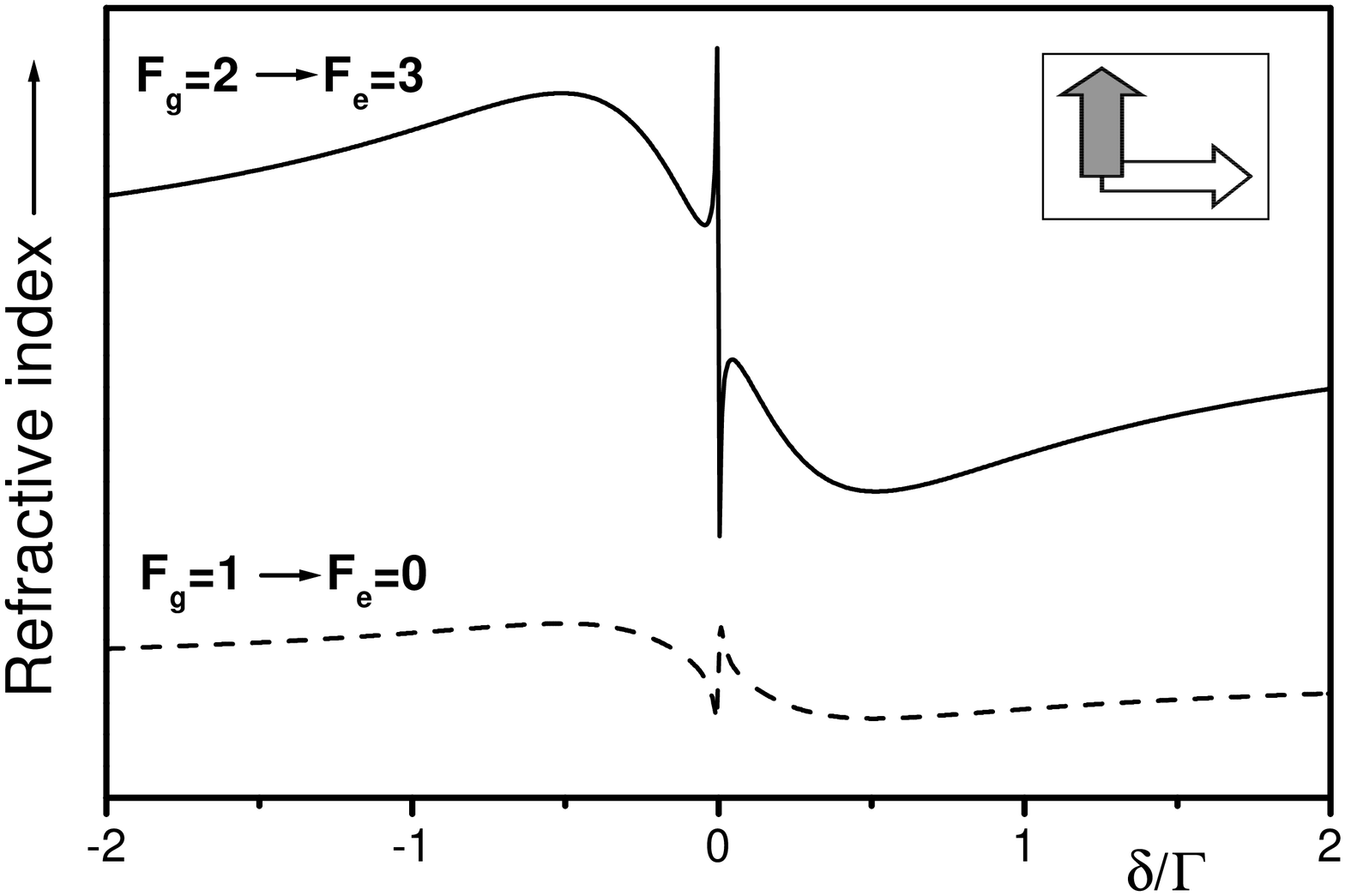,width=3.5in}}
\end{center}
\caption{Refractive index spectra for the probe field as a function of the
frequency offset $\delta $ for two closed atomic transitions. $B=0$, $\Omega
_1/\Gamma =0.4,\ \gamma /\Gamma =0.001$. The pump and probe polarizations
are linear and orthogonal. The vertical positions of the traces are
arbitrary. The vertical scale is the same for both curves.}
\label{steeprefrac}
\end{figure}

The sharp variations in the absorption corresponding to EIT or EIA
resonances are accompanied by modifications of the refractive index. Fig. 
\ref{steeprefrac} shows the calculated refractive index variation for the
probe field for two different closed transitions with linear and orthogonal
pump and probe polarizations. Very steep normal and anomalous dispersion
takes place for EIT and EIA respectively\cite{MESCHEDE} corresponding to
small absolute values of the group velocity which may be negative in the
case of EIA\cite{AKULSHIN}.

\subsubsection{Spectra in the presence of a magnetic field.}

\begin{figure}[tbp]
\begin{center}
\mbox{\epsfig{file=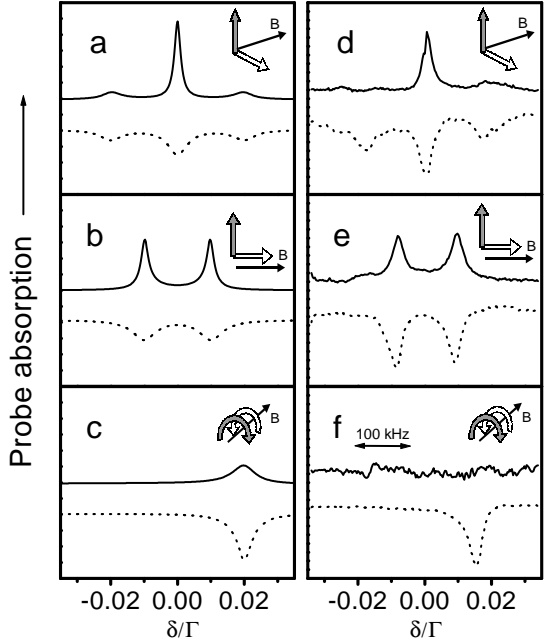,width=3.5in}}
\end{center}
\caption{{\bf a-c}: Calculated probe absorption spectra under the presence
of a magnetic field ($\beta _gB/\Gamma =0.01,\ \Omega _1/\Gamma =0.3,\
\gamma /\Gamma =0.001$) for the transitions $F_g=3\rightarrow F_e=4$ (solid)
and $F_g=2\rightarrow F_e=1$ (dashed). The vertical positions of the traces
are arbitrary. The vertical scale is the same for all curves. {\bf d-f}:
Measured probe absorption spectra with the laser tuned to the transitions $%
5S_{1/2}(F_g=3)$ $\rightarrow 5P_{3/2}$(solid) and $5S_{1/2}(F_g=2)%
\rightarrow 5P_{3/2}$ (dashed) of $^{85}$Rb under the presence of a magnetic
field ($\beta _gB/2\pi \approx 50\ kHz,\ \Gamma /2\pi =5.9\ MHz$). The
vertical positions of the traces are arbitrary. The vertical scale is the
same for all curves.}
\label{calcwb}
\end{figure}

The presence of a magnetic field removes the degeneracy of the atomic level.
As a consequence, the narrow resonances corresponding to EIT or EIA split
into several components. The position of each component is given by a
resonance condition for a Raman transition between two ground-state Zeeman
sublevels involving one pump photon and one probe photon. The selection
rules determining the number of resonances are the same for EIT or EIA.
However the spectra obtained under the same polarizations and intensities of
the pump and probe fields, present some qualitative and quantitative
differences in addition to the sign inversion. Fig. \ref{calcwb} present the
comparison of the EIT\ and EIA spectra calculated for the closed transitions 
$F_g=3\rightarrow F_e=4$ and $F_g=2\rightarrow F_e=1$. It is interesting to
point out that whenever it is allowed by the Raman selection rules, the line
corresponding to transitions from Zeeman sublevels to themselves is present
in the spectra. This is the case of Fig. \ref{calcwb}a. The corresponding
peak occurs at zero pump-probe frequency offset independently of the value
of the magnetic field. Also for some combinations of the pump and probe
polarizations (for instance $\sigma ^{+}$ and $\sigma ^{-}$ respectively)
there is a unique resonance whose position depends on the value of the
magnetic field (Fig. \ref{calcwb}c).

\subsection{Four-wave-mixing spectra.}

The importance of level degeneracy on FWM generation was long ago
appreciated. It plays an essential role in connection with the study of
collision-induced resonances\cite{ROTHBERG,BERMAN-STEEL,BERMAN}. Recently,
FWM in a degenerate two-level system was used for the construction of a
phase conjugate resonator\cite{HEMMER}. We analyze now the predictions of
our model concerning the generation of a new field by FWM according to Eq. 
\ref{fwm}.

\begin{figure}[tbp]
\begin{center}
\mbox{\epsfig{file=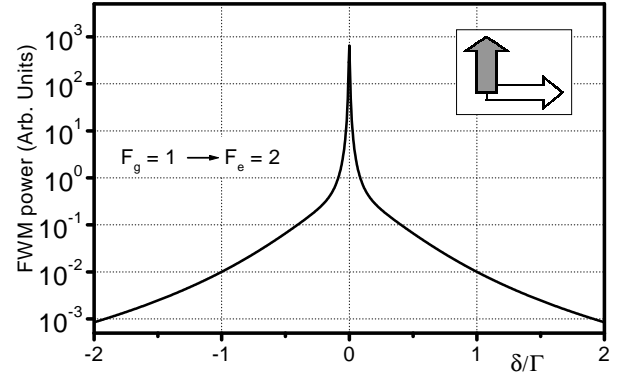,width=3.5in}}
\end{center}
\caption{Power of FWM emission as a function of the frequency offset $\delta 
$ for the closed $F_g=1\rightarrow F_e=2$ transition ($B=0,\ \gamma /\Gamma
=0.001)$. The pump and probe polarizations are linear and orthogonal.}
\label{fwmspeclog}
\end{figure}

When the conditions for coherence resonances are satisfied a large increase
in the intensity of the field generated at frequency $2\omega _1-\omega _2$
occurs. The generated FWM power as a function of $\delta $ is presented in
Fig. \ref{fwmspeclog} in the case of orthogonal and linear pump and probe
polarizations and no magnetic field (notice the vertical logarithmic scale).
As seen the coherence resonance peak of FWM can be several orders of
magnitude larger than the non-resonant background of width $\sim \Gamma $.
The high contrast is a consequence of the triple resonant interaction of the
three fields involved in the nonlinear mixing process \cite{SHEN}.

\begin{figure}[tbp]
\begin{center}
\mbox{\epsfig{file=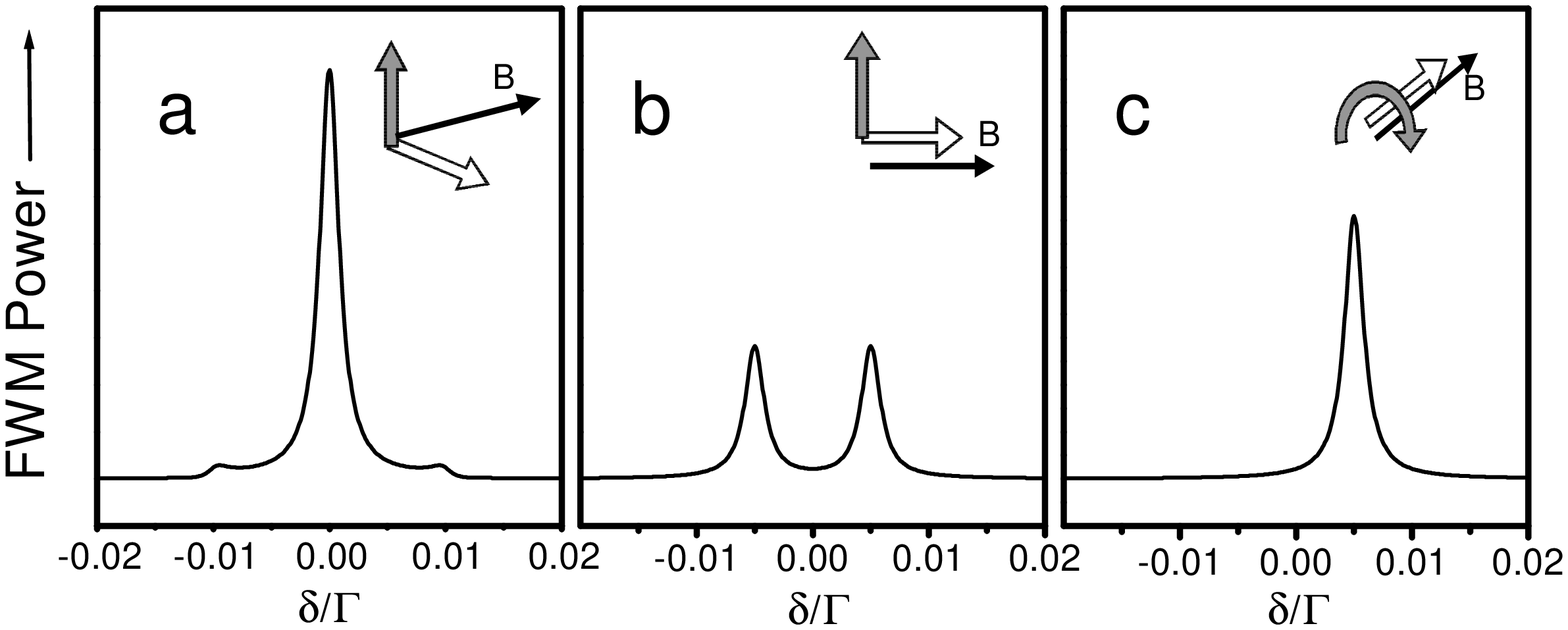,width=3.5in}}
\end{center}
\caption{Power of FWM emission as a function of the frequency offset $\delta 
$ for the closed $F_g=1\rightarrow F_e=2$ transition under the presence of a
magnetic field ($\beta _gB/\Gamma =0.005,\ \gamma /\Gamma =0.001$). The pump
and probe polarizations are: $a)$ linear, orthogonal and transverse respect
to $\vec{B}$. $b)$ linear, orthogonal and probe along $\vec{B}$. $c)$ pump
circular transverse, probe linear along $\vec{B}$.}
\label{fwmspeclin}
\end{figure}

Under the presence of a magnetic field the FWM spectra splits into several
peaks. Fig. \ref{fwmspeclin} shows some examples of calculated spectra
obtained for different pump and probe polarizations. The selection rules
governing resonant FWM result from two simultaneous requirements on the pump
and probe polarizations: $a)$ A three-photon transition from the ground to
the excited level involving the absorption of two pump photons and the
emission of one probe photon should be permitted. $b)$ An electric dipole
transition should be allowed between the initial (ground) and the final
(excited) sublevels connected by the three-photon processes. These rules are
different from the rules corresponding to probe absorption. For instance, no
coherence resonance in FWM occurs if the pump and probe polarizations are
circular and opposite. However, absorption resonances are generally
observable in this case.

To our knowledge, the results presented here constitutes the first detailed
calculation of coherence resonances spectra in FWM involving Zeeman
sublevels in degenerate two-level system.

\subsection{Population modulation spectra.}

The modulation of the excited state population results in modulation of the
emitted fluorescence according to Eq.\ref{acfluor}. This modulation has a
similar origin than that discussed in the context of CPT in a driven $%
\Lambda $ system \cite{GRISHANIN}. As mentioned in \cite{GRISHANIN} the
modulation in the fluorescence is intimately connected to FWM. In the
context of the present paper, this connection is the natural consequence of
the coupling between $\sigma _{ge}^{+}$, $\sigma _{eg}^{+}$, $\sigma
_{gg}^{+}$ and $\sigma _{ee}^{+}$ described by Eq.\ref{eqprobe}.

\begin{figure}[tbp]
\begin{center}
\mbox{\epsfig{file=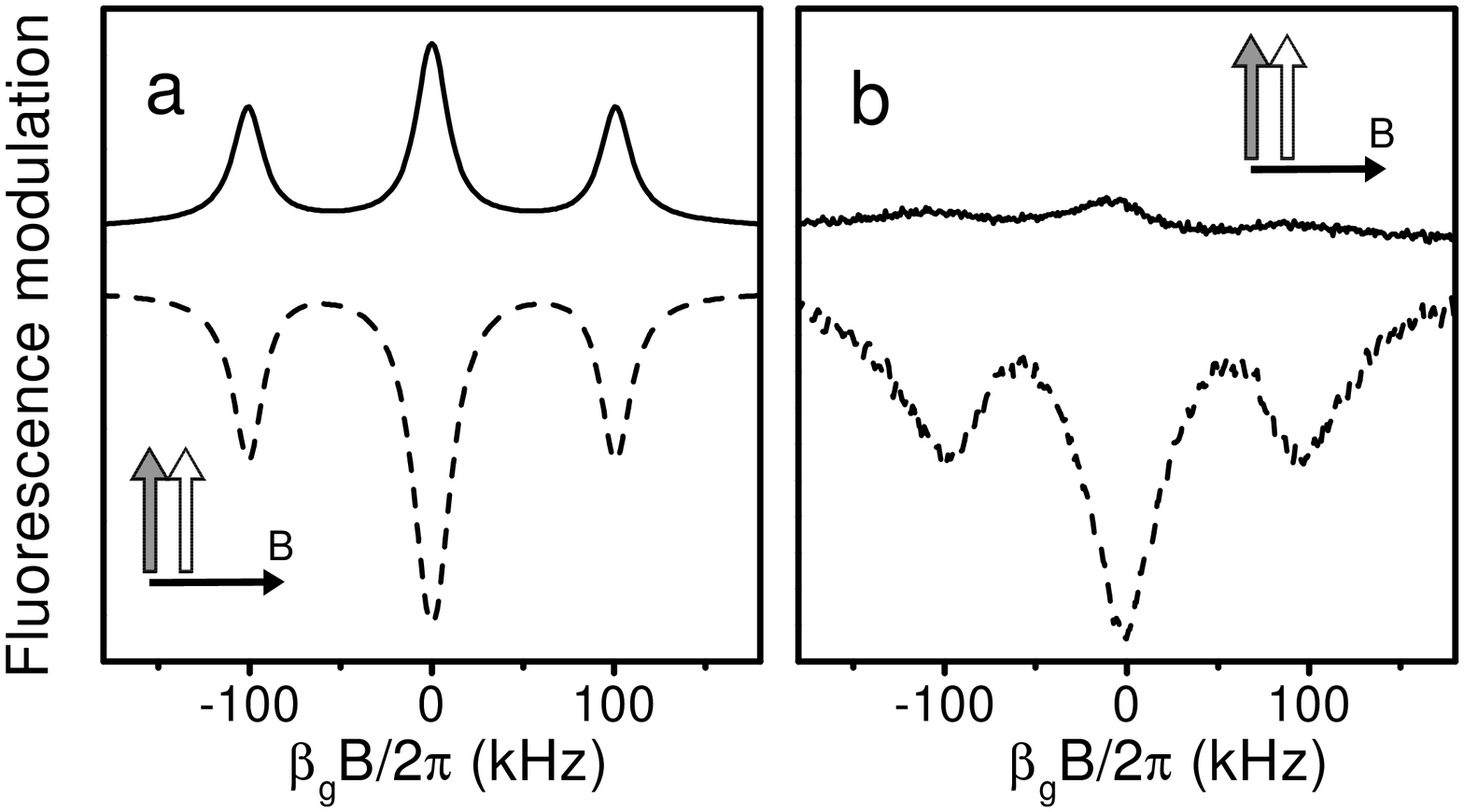,width=3.5in}}
\end{center}
\caption{Fluorescence modulation at the frequency of the pump to probe
offset $\delta /2\pi =200\ kHz$ as a function of magnetic field. $a)$
Calculated signal for the closed transitions $F_g=3\rightarrow F_e=4$
(solid) and $F_g=2\rightarrow F_e=1$ (dashed) with $\Omega _1/\Gamma =0.3,\
\gamma /\Gamma =0.001,\ \Gamma /2\pi =5.9\ MHz$). $b)$ Measured signal with
the laser tuned to the transitions $5S_{1/2}(F_g=3)$ $\rightarrow 5P_{3/2}$%
(solid) and $5S_{1/2}(F_g=2)\rightarrow 5P_{3/2}$ (dashed) of $^{85}$Rb. The
pump and probe polarizations are linear, parallel and transverse with
respect to $\vec{B}$. The vertical positions of the traces are arbitrary.}
\label{pulsation}
\end{figure}

Population modulation is due to the coupling of the optical coherence
induced by one of the fields between two Zeeman sublevels in the ground and
the excited state with the second field. In consequence, it occurs when the
two fields couple the same pair of excited and ground Zeeman sublevels. If
the polarizations of the two fields are the same and correspond to a proper
polarization with respect to the magnetic field orientation ($\sigma ^{+}$, $%
\sigma ^{-}$ or $\pi $), the modulated fluorescence present no resonance
narrower than $\Gamma $ as a function of the pump to probe frequency offset.
However, interesting interference effects, resulting in coherence resonances
(linewidth determined by $\gamma $), occur for other pump and probe
polarizations. Fig. \ref{pulsation}(a,b) shows the calculated modulated
component of the total fluorescence for two different transitions as a
function of the magnetic field for fixed pump-probe frequency offset and the
same transverse linear polarization. Three coherence resonances are present
in this case. The central one corresponds to zero magnetic field. The
positions of the two lateral peaks correspond to the conditions for resonant
Raman transitions between Zeeman sublevels. The signs of the resonances are
opposite in the case of transitions corresponding to EIA (Fig.\ref{pulsation}%
a) or EIT (Fig.\ref{pulsation}b). Notice the larger relative variation of
the pulsation amplitude in the case of EIT.

\subsection{Oscillating magnetic dipole spectra.}

The simultaneous interaction of an atomic system with two fields can result
in the driving of Zeeman coherence in the ground state. This coherence can
in turn be responsible for electromagnetic emission \cite{VANIER}. To
illustrate this phenomenon, in the case of degenerate two-level system, we
analyze the predictions of our model concerning the induced ground-state
magnetic dipole oscillating at frequency $\delta $ (Eq.\ref{mag_dip}).

\begin{figure}[tbp]
\begin{center}
\mbox{\epsfig{file=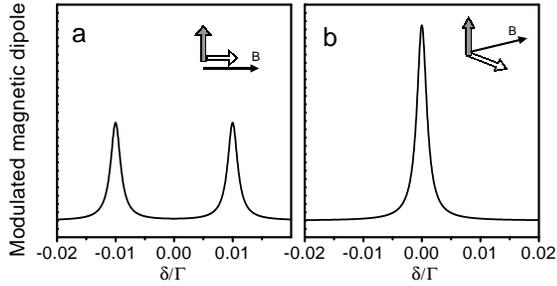,width=3.5in}}
\end{center}
\caption{Calculated modulus of the ground state magnetic dipole modulated at
frequency $\delta $ for $\beta _gB/\Gamma =0.01,\ \Omega _1/\Gamma =0.01,\
\gamma /\Gamma =0.001$. The pump and probe polarizations are: $a)$ pump
linear transverse, probe linear along $\vec{B}$. $b)$ pump and probe linear
orthogonal and transverse respect to $\vec{B}$.}
\label{orient}
\end{figure}

The selection rules governing the oscillating magnetic dipole spectra differ
from those of the previously analyzed processes. Some examples of the
predicted oscillating magnetic dipole as a function of the magnetic field
for fixed pump to probe frequency offset are shown in Fig.\ref{orient}.
Coherence resonances in the oscillating dipole modulus take place when the
two fields couple adjacent ground state Zeeman sublevels. This is the case
of Fig.\ref{orient}$a$ corresponding to $\pi $ polarized pump and a
transverse and linearly polarized probe. The coherence resonance illustrated
in Fig.\ref{orient}$b$ corresponding to transverse, orthogonal linearly
polarized pump and probe is somehow different since its position ($B=0$)
does not correspond to a two photon process between different Zeeman
sublevels. The resonance can be understood, in this case, as the result of
interference between two non resonant two-photon processes. In every case
the resonance width are determined by the relaxation rate $\gamma $.

\section{Experiments.}

\subsection{Setup.}

The experiments were performed on the D$_2$ line of $^{85}$Rb. The
experimental setup is similar to the one previously presented\cite
{LEZAMA,LEZAMA2}. We remind here the principal features. The pump and probe
fields were generated from the output of a unique extended cavity diode
laser. This laser was locked, with the help of an external servo loop, to a
saturated absorption line obtained from a reference Rb vapor cell. The
frequency uncertainty of the locked laser is less than 1 MHz over several
minutes. Most of the power of the laser beam was used as the pump field. The
probe field was generated by frequency shifting a fraction of the laser beam
with two consecutive acousto-optic modulators, one of them driven by a
variable RF source. In this way two mutually coherent pump and probe waves
were generated with tunable frequency offset $\delta $. The polarization of
the two fields were independently controlled. Perfect overlap between the
two fields was achieved by propagating them along a 50 cm single mode
optical fiber that did not significantly modify the polarizations. The light
was sent through a 2 cm long vapor cell. The power of the pump and probe at
the atomic sample were $0.8\ mW$ and $50\ \mu W$ respectively. The atomic
cell was placed within Helmholtz coils for magnetic field control. For the
probe absorption measurements, the pump beam was blocked by a polarizer,
(extinction ratio larger than 200) while the probe intensity was detected
with a photodiode. To enhance sensitivity and signal to noise ratio, a
two-frequency lock-in detection technique was used \cite{DEMTRODER}. For
this, the pump and probe fields were chopped at the frequencies $f_1$ and $%
f_2$ (around 1 kHz) respectively and the photodiode output current was
analyzed at frequency $f_1+f_2$ by a lock-in amplifier. By this means only
the non-linear component of the probe transmission, dependent on the pump
and probe intensities, was detected.

For the measurement of the excited state population modulation, the atomic
fluorescence was collected with a large area photodiode (1cm diameter)
situated close to the vapor cell. The photodiode current was sent to an RF
frequency analyzer operating in the zero span mode, that is measuring the RF
input amplitude at a fixed frequency set equal to the probe to pump
frequency offset $\delta $. A fixed value of $\delta =200\ kHz$ was used
while the longitudinal magnetic field at the cell was scanned over a few
hundred milligauss around the zero magnetic field. During these measurements
the cell was surrounded by a $\mu -$metal shield reducing the ambient
magnetic field to less than $10\ mG$. The scanning of the longitudinal
magnetic field was accomplished with coils placed inside the magnetic shield.

\subsection{Results.}

\subsubsection{Probe absorption.}

Fig. \ref{calcwb}(d-f) shows several examples of probe absorption spectra
obtained for different orientations of the magnetic field and the optical
fields polarizations. In each figure division, two spectra obtained from the
excitation of the lower and the upper ground-state hyperfine levels of $%
^{85} $Rb under the same optical fields intensities and polarizations are
presented. When the lower ground-state hyperfine level is addressed EIT
resonances are observed while in the case of the upper ground-state
hyperfine level EIA occurs. The agreement between the predictions of the
model and the observations is quite satisfactory. This agreement may seem
rather surprising since in the vapor cell, due to the velocity distribution,
three different atomic transitions, one closed and two open contribute to
the signal in each case. On all open transitions as well as on the closed $%
F_g=2\rightarrow F_e=1$ transition EIT occurs. Only on the $F_g=3\rightarrow
F_e=4$ transition EIA takes place \cite{LEZAMA2}. However, due to optical
pumping the absorption signal is essentially determined by the closed
transitions resulting in the qualitative agreement with the theoretical
prediction. Some differences are nevertheless observed: The relative
strength of the EIA and EIT transitions are not the same than in the
calculation and the small EIA resonance predicted in Fig.\ref{calcwb}c is
not visible in the experiment (Fig.\ref{calcwb}f). We have verified that a
closer agreement between the experimental observation and theory can be
achieved by taking into account the excited state hyperfine structure and
the atomic velocity distribution according to the procedure described in 
\cite{LEZAMA2}.

\subsubsection{Modulated fluorescence.}

The measured signal corresponding to the fluorescence modulated at frequency 
$\delta =200kHz$ is presented in Fig.\ref{pulsation}(c,d). The traces are
composed of a constant background over which narrow resonances are visible.
The modulated fluorescence is of the order of 10$^{-3}$ times the total
fluorescence. The contrast of the coherence resonances with respect to the
constant background is larger in the case of the lower ground state
hyperfine level ($\sim 20\%$) compared to that of the upper ground state
hyperfine level ($\sim 5\%$). The width of the peaks are determined by
magnetic field inhomogeneities.

The main features in the observed curves are in good qualitative agreement
with the predictions based on the theoretical model valid for atoms at rest.
This indicates that, as in the case of absorption, the spectra are dominated
by the corresponding cycling transition and that the influence of the
Doppler effect on the coherence resonances positions is not essential. More
elaborate calculations including velocity integration and summation over
excited state hyperfine levels provide closer agreement with the
observations \cite{METAL}. The results reported here constitute, to our
knowledge, the first demonstration of the observation of coherence
resonances through the analysis of fluorescence modulation.

\subsection{Conclusions.}

The response of a degenerate two-level atomic system to the simultaneous
presence of a pump and a probe field has been examined with the help of a
theoretical model allowing the numerical calculation of the atomic response
in a large variety of situations. Based on this model, the manifestations of
the coherent nature of the atom-field interaction on different observables
were analyzed and the corresponding spectra calculated. Most of these
observables present narrow coherence resonances whose width is essentially
determined by the ground-state relaxation. The dependence of the coherence
resonances spectra on the atomic transition, the optical fields
polarizations and magnetic field was investigated. Two experimental
observations are compared with the predictions of the model: the probe
absorption spectra in the presence of a magnetic field and the modulated
fluorescence dependence on magnetic field.

The results reported in this paper illustrate the richness of the coherent
response of degenerate two level systems. Further study of coherence
resonances in degenerate two-level systems should broaden our present
understanding of coherent processes. In addition, it may provide the base
for interesting applications such as magnetic field measurement\cite{LEE} or
refractive index manipulation\cite{AKULSHIN}.

\subsection{Acknowledgments.}

The author wish to thank D. Bloch and M. Ducloy for stimulating discussions.
This work was supported by CONICYT, CSIC and PEDECIBA (Uruguayan agencies)
and ECOS (France).


\end{document}